\documentclass[journal,comsoc]{IEEEtran}

\usepackage{epsfig}
\usepackage{amsmath,amssymb,bm}
\usepackage{array}
\usepackage{enumitem}
\usepackage{multicol}
\usepackage{graphicx}
\usepackage{color,soul}

\begin{document}
\pagenumbering{gobble}

\title{Securing Heterogeneous IoT with Intelligent DDoS Attack Behavior Learning}

\author{Nhu-Ngoc~Dao,
        Trung~V.~Phan,
        Umar~Sa'ad,
        Joongheon~Kim,
        Thomas~Bauschert,
        and Sungrae~Cho \\
        
    \thanks{N.-N. Dao, U. Sa'ad, J. Kim, and S. Cho are with Chung-Ang University, 
            School of Computer Science and Engineering, 
            Seoul, Republic of Korea.}
    \thanks{T.V. Phan and T. Bauschert are with Chemnitz University of Technology, Faculty of Electrical Engineering and Information Technology, Chemnitz, Germany.}
    \thanks{Corresponding author: S. Cho (srcho@cau.ac.kr)}
    \thanks{\hl{“This work has been submitted to the IEEE for possible publication. Copyright may be transferred without notice, after which this version may no longer be accessible.”}}
}

\maketitle

\begin{abstract}
The rapid increase of diverse Internet of things (IoT) services and devices has raised numerous challenges in terms of connectivity, computation, and security, which networks must face in order to provide satisfactory support. This has led to networks evolving into heterogeneous IoT networking infrastructures characterized by multiple access technologies and mobile edge computing (MEC) capabilities. The heterogeneity of the networks, devices, and services introduces serious vulnerabilities to security attacks, especially distributed denial-of-service (DDoS) attacks, which exploit massive IoT devices to exhaust both network and victim resources. As such, this study proposes MECshield, a localized DDoS prevention framework leveraging MEC power to deploy multiple smart filters at the edge of relevant attack-source/destination networks. The cooperation among the smart filters is supervised by a central controller. The central controller localizes each smart filter by feeding appropriate training parameters into its self-organizing map (SOM) component, based on the attacking behavior. The performance of the MECshield framework is verified using three typical IoT traffic scenarios. The numerical results reveal that MECshield outperforms existing solutions.
\end{abstract}


\IEEEpeerreviewmaketitle

\section{Introduction}
In a recent report by Gartner~\cite{gartner2017report}, 20.4 billion Internet of things (IoT) devices are expected to be in use by 2020, an increase of 219\% from 2016. These devices have become popular in whole market segments, including consumer applications, cross-industry business, and vertical-specific industry. 

This IoTization boom poses several challenges for IoT networks, including access connectivity, offloading computation, and security. In particular, lightweight IoT devices, which are typically characterized by low computing power, are exploited by attackers to generate flooding traffic in distributed denial-of-service (DDoS) attacks. For instance, a swarm of IoT devices, hijacked by Mirai malware, generated about 1 Tbps of DDoS traffic to a French webhost in September 2016 \cite{kolias2017ddos}. These vulnerabilities pose serious challenges to IoT systems. 

Consequently, networking infrastructures are evolving into heterogeneous IoT (H-IoT) networks, which are characterized by diverse IoT devices, multiple access technologies, and powerful computation provided at the edge tier, referred to as mobile edge computing (MEC) \cite{sun2016edgeiot}. Along with the softwarization capability of emerging technologies, such as software-defined network function virtualization (SDNFV), MEC-enabled H-IoT networks support a promising framework for H-IoT security. Security-as-a-service (SaaS) can be deployed with the support of MEC technology to prevent internal/external attacks from/to the H-IoT networks.

Based on the above analysis, we propose a DDoS prevention framework, called \textit{MECshield}. MECshield utilizes MEC technology to provide localized smart filters against malicious threats at the edge of relevant attack-source/destination networks. A central controller orchestrates the cooperation among the smart filters through policy dispatches. The policies are generated based on attack-behavior analyses conducted at an attack analyzer in the central controller, before being dispatched to the corresponding smart filters at the attack-source sites. In each smart filter, a self-organizing map (SOM) component \cite{kohonen2013essentials} is trained simultaneously using local traffic, under the supervision of the dispatched policy. The trained SOM detects malicious IoT traffic by matching the traffic features into the SOM map in order to identify whether it represents a DDoS attack.

Accordingly, the MECshield framework is characterized by:
\begin{itemize}
    \item \textit{Cooperation,} where the central controller distributes the features of malicious traffic analyzed at the victim sites to the corresponding smart SOM filters at the attack-source sites via policy dispatches. 
    \item \textit{Distribution,} where malicious threats are handled simultaneously at multiple attack-source sites and at the attack-destination sites. In other words, bottleneck problems, which are a crucial issue in DDoS attacks, are mitigated.
    \item \textit{Local adaptation,} where each smart SOM filter is trained using local traffic features collected from the H-IoT on which the smart SOM filter has been deployed.
\end{itemize}

\begin{figure*}[!t]
    \centering
    \includegraphics[width=0.8\textwidth]{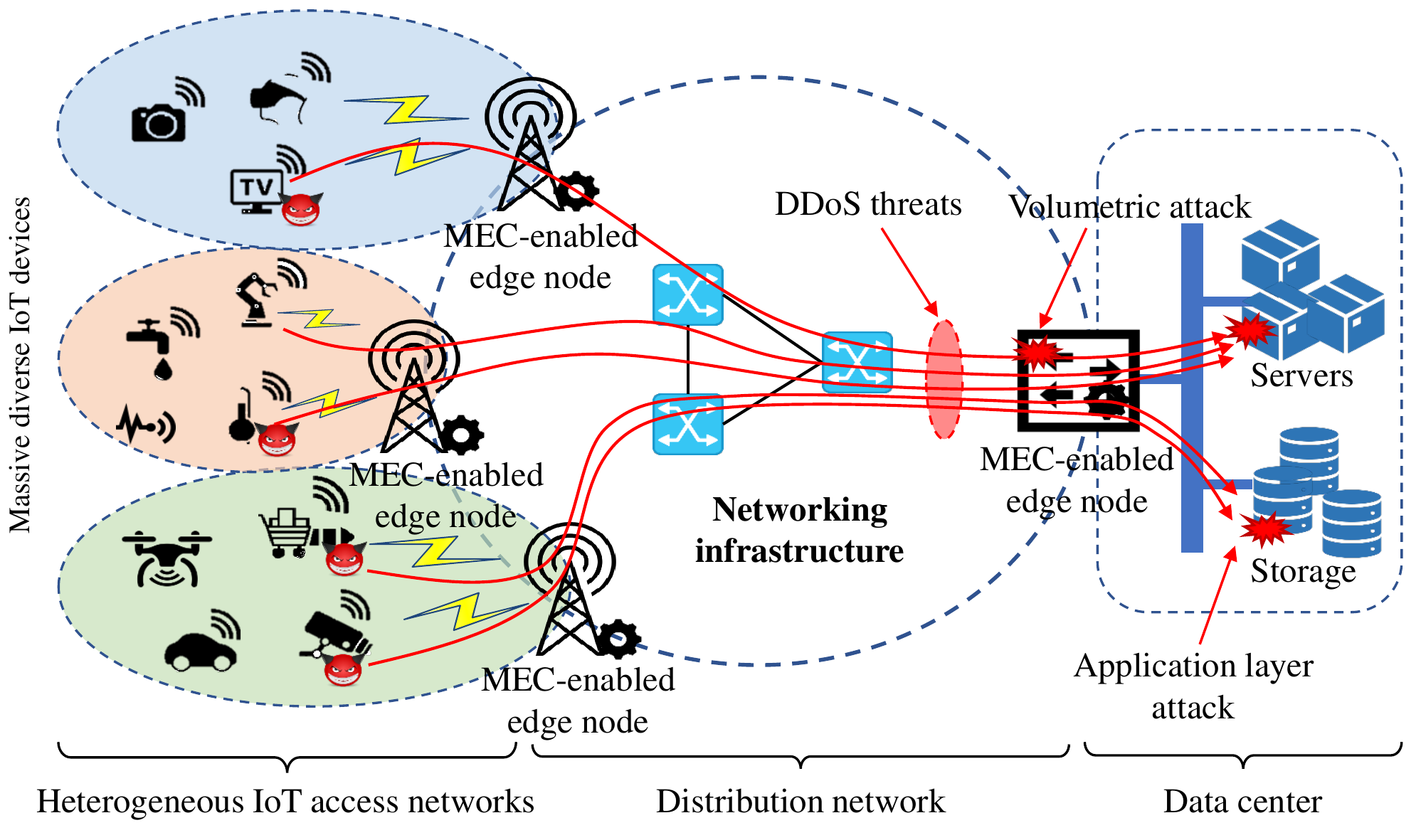}
    \caption{Scenarios of DDoS attacks in MEC-enabled H-IoT networks.}
    \label{fig:network}
\end{figure*}

There are three benefits of the MECshield framework:
\begin{itemize}
    \item \textit{Intermediate traffic reduction:} Since smart SOM filters are deployed in front of the attack source, malicious traffic generated by H-IoT devices is typically blocked at the edge. 
    \item \textit{Detection and accuracy improvement:} This is the result of two main features of MECshield: (i) continuous localized training, and (ii) implementation in proximity to the attack source and destination. The first feature adapts the smart SOM filter to time/position-varying H-IoT traffic. Meanwhile, the second feature provides two protection layers against malicious traffic.
    \item \textit{Prevention performance increase:} Owing to its distributed architecture, MECshield avoids the bottleneck problem typical in DDoS attacks. Moreover, the two protection layers at the attack-source/destination sites are supervised by the central controller for traffic handling orchestration.
\end{itemize}

\section{MEC-enabled Heterogeneous IoT Networks}
This section describes the features of MEC-enabled H-IoT networks for determining their vulnerabilities (\textit{V1}, \textit{V2}, and \textit{V3}) and resistance (\textit{R1}, \textit{R2}, and \textit{R3}) to DDoS attacks; see Fig.~\ref{fig:network} for the reference model. MEC-enabled H-IoT networks consist of various local IoT networks distributed at the edge and interconnected via the core infrastructure. The following features are derived from two distinguishing facets of the network: the heterogeneity of IoT devices, and external computation served by MEC. 

\textit{Power-constrained devices (V1):} Although IoT does not exclude high-power devices, those with constrained power in terms of computing resources and memory typically occupy the dominant position \cite{lin2017survey}. Owing to their lack of computational power, these IoT devices may not support complex and evolving security algorithms, such as effective encryption for data transfer and endpoint protection against local security attacks. Furthermore, the weak security implemented on these devices (e.g., default and hard-coded passwords) means exploiting and recruiting them into botnets and injecting different types of malware are trivial tasks for even unskilled attackers (\textit{script kiddies}). 

\textit{Massive connections (V2):} Billions of connected IoT devices generate massive volumes of data. This is an important ingredient for effective DDoS attacks. The traffic is usually generated from many constrained H-IoT devices. However, the same amount of traffic might also be generated from fewer powerful devices in other networks. These factors make H-IoT traffic containing malicious DDoS flows more difficult to handle than other network traffic.

\textit{Heterogeneous group-specific traffic (V3):} H-IoT traffic is considered  heterogeneous from a macro perspective, but group-specific from the perspective of each local network \cite{alam2017data}. In particular, IoT devices serving individual applications may be separately connected in different virtual local area networks (VLANs), which can be managed at the edge of the H-IoT networks. In general, each IoT application transfers data in its own way. As such, the generated traffic can be identified via a tuple of flow parameters, such as protocols, ports, transmission rates, and transmission contiguity. From a security viewpoint, the aggregated traffic at the attack-destination site is classified into a heterogeneous category, while the outgoing traffic from the attack-source sites is divided into a group-specific category. This classification is meaningful for the adaptive development of security strategies against malicious traffic.

\textit{Edge cloudization (R1):} MEC technology provides cloudization capabilities at the edge of the H-IoT network. This is characterized by ultra-low execution latency and context-aware computation \cite{hu2015mobile}. This environment enables services such as resource scheduling, and security protection to be scalably deployed in proximity to the IoT devices. Therefore, comprehensive DDoS prevention, facilitated by the MEC, can be implemented in collaboration with advanced techniques such as machine learning and big data mining in a local context.

\textit{Service execution offloading (R2):} Edge cloudization has enabled the increasingly popular service execution offloading in H-IoT. While lightweight IoT devices lack the powerful computation capability necessary for the timely execution of complex services, the networks are equipped with sufficient computational resources to provide tailored service execution on demand. This trend has resulted in traffic behavior that prioritizes local processing at the edge, rather than on Internet servers. Tracking this traffic behavior of each local network might help to detect security threats when abnormal traffic behavior occurs.

\textit{Contextual information fusion (R3):} Although the traffic properties can be distinguished among local IoT networks, applications running at the edge may need to merge contextual IoT data to obtain comprehensive information. The relationships among contextual IoT data can be considered a criterion for abnormal traffic detection when individual partners defect \cite{aleroud2017contextual}. For instance, standard images are transferred from cameras to a surveillance system during the day, while thermographic images and motion detection signals are more useful at night. The traffic is considered abnormal when, for example, thermographic images are sent during the day, or standard images are sent at night.

\section{Security Problem Statement}

This section analyzes the adversary models of two DDoS attack scenarios (a volumetric DDoS attack and an application layer attack) in terms of the attack objectives, initial capabilities, and process.

\textbf{Objective:}
The objectives of the scenarios are as follows: 
\begin{enumerate}[label=(\roman*)]
\item\textit{Scenario 1} -- A volumetric DDoS attack on the infrastructure between users on the Internet and a data center. The objective is to send lots of bogus traffic generated from compromised H-IoT devices so that total malicious traffic size exceeds the capacity of the network. This is the most prevalent type of DDoS attack, constituting 73\% of all DDoS attacks experienced in 2016 \cite{network2017worldwide}.
\item\textit{Scenario 2} -- An application layer DDoS attack on a server. The objective is to flood the server with seemingly legitimate, but bogus requests in order to exhaust the ability of the application to serve legitimate users. This is a more sophisticated type of DDoS attack, and is difficult to detect because the attack traffic is not easily distinguishable from benign traffic. The number of such attacks increased by 95\% between 2015 and 2016. \cite{network2017worldwide}.
\end{enumerate}

\textbf{Initial capabilities:}
In order to execute the attacks, we assume the adversary has the following capabilities:
\begin{itemize}
    \item\textit{Botnet:} Access to a group of compromised IoT devices (H-IoT botnet). The adversary may be the owner of the botnet (botmaster), or may have access to it through a third party (e.g., a DDoS-for-hire service).
    \item\textit{Command and Control (C2):} A command and control infrastructure (C2), which is used to control the compromised devices and, possibly to recruit additional devices. 
    \item\textit{System Knowledge:} Some knowledge about the victim, such as IP addresses, domain names, existing vulnerabilities, and so on.
    \item\textit{Amplifiers:} Poorly configured network services (e.g., Open DNS resolver), which the attacker can exploit to increase the volume of the generated botnet traffic. This capability is crucial for the attack in scenario 1.
    \item\textit{IP Spoofing:} Ability to spoof the source IP address of the botnet traffic. This capability \textit{reflects} the amplified botnet traffic by sending it to the victim rather than the real source.
\end{itemize}

\textbf{Attack process:}
The attack process in each scenario is described as follows:
\subsubsection{Scenario 1}
\begin{itemize}
    \item\textit{Botnet Activation:} The attacker uses a controller to send commands to the H-IoT botnet. The instructions may include the victim’s IP address, attack rate, target service (DNS, NTP, and SSDP etc.).
    \item\textit{Traffic Generation:} The botnet is used to generate traffic using the above parameters.
    \item\textit{Amplification:} Some UDP-based network protocols have a high bandwidth amplification factor, which simply means they return very large responses for much smaller requests. For example, DNS has an amplification factor of 28 to 54, NTP has a factor of 556.9, and SSDP has a factor of 30.8 \cite{leverett2017towards}. This property is exploited by attackers in volumetric DDoS attacks, in which a large H-IoT botnet is used to send requests to these services in order to generate an enormous amount of traffic as a response.
    \item\textit{Reflection:} The source IP address of the botnet packets is spoofed and replaced with the IP address of the victim. Therefore, the amplified traffic is sent to the victim rather than to the attacker.
    \item\textit{Network Disruption/Degradation:} The network capacity is eventually exceeded by the amplified and reflected traffic, thereby degrading or disrupting the operations of the network.
\end{itemize}

\subsubsection{Scenario 2}
\begin{itemize}
    \item\textit{Botnet Activation:} This process is the same as that in scenario 1.
    \item\textit{Traffic Generation:} At this stage, traffic is generated from each compromised device in the H-IoT botnet. The intent of the attacker is not easily discernible because the traffic conforms to all protocols.
    \item\textit{Flooding:} At this stage, the attacker floods the server with requests from each compromised device in the H-IoT botnet. There are three types of application layer flooding attacks: session flooding, where each device sends sessions at higher rates than those of non-malicious users; request flooding, where each attack session involves sending higher requests than those of non-malicious users; and an asymmetric attack, where each attack session contains requests with much higher workloads than those of non-malicious sessions.
    \item\textit{Service Disruption/Degradation:} The capacity of the server to respond to user requests is eventually exceeded, thus making the server unavailable.
\end{itemize}

\begin{figure*}[!t]
    \centering
    \includegraphics[width=0.8\textwidth]{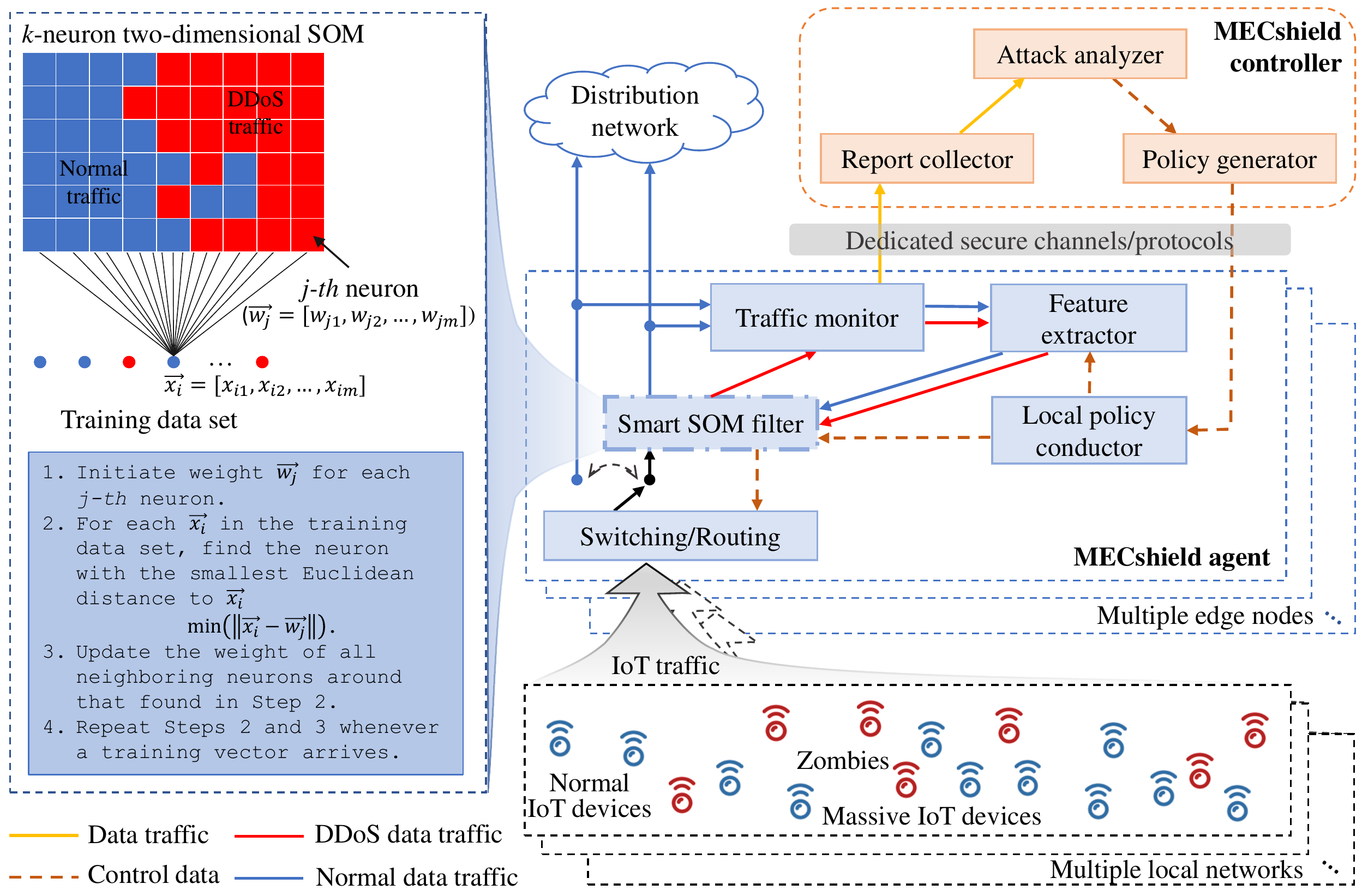}
    \caption{MECshield framework for heterogeneous IoT.}
    \label{fig:architecture}
\end{figure*}

\section{MECshield Framework}
Based on the previously mentioned security problems, we propose MECshield, a novel DDoS prevention framework.

\subsection{Design Rationale}
The rationale behind the MECshield framework design includes (i) utilizing \textit{MEC technology} to provide \textit{distributed} security agents in front of the attack-source/destination sites; (ii) \textit{well-adaptation} to local traffic of the SOM filter at each agent with purpose of abnormal detection improvement; and (iii) \textit{cooperation} among the agents, supervised by a \textit{central controller}. The smart SOM filter in each MECshield agent simultaneously classifies outgoing traffic from IoT devices and is trained by local traffic, monitored in real-time.

\subsection{Self-Organizing Map algorithm}

The SOM algorithm is one of the most effective unsupervised learning solutions in Artificial Neural Networks, which converts a higher-dimensional input space into a lower-dimensional representation called a SOM as illustrated on the left side of Figure \ref{fig:architecture}. The SOM classifies a new input vector based on two main modes: training and mapping. The former builds the map using input samples, while the latter classifies new input vectors by finding a winning neuron which has the smallest Euclidean distance in the map. In this work, we utilize a \textit{k}-neuron two-dimensional SOM, where the \textit{j-th} neuron has a weight vector $\overrightarrow{w_j}=\left [ w_{j1}, w_{j2}, \ldots, w_{jm} \right ]$, within a dimension \textit{m}, equal to the number of considered traffic features. Initially, the weight vector of each neuron in the SOM is generated randomly and the feature values in a training vector is always formed into an $[0.0,1.0]$ range. Thereafter, whenever an \textit{i-th} training vector $\overrightarrow{x_i}$ = $\left [ x_{j1}, x_{j2}, \ldots, x_{jm} \right ]$ arrives, the winning neuron $\overrightarrow{w^*_i}$ is selected by finding the minimum Euclidean distance between the training vector and the neurons, as follows:
\begin{equation}
    \overrightarrow{w^*_i} = \underset{\forall \overrightarrow{w_j} \in W}{\min}\left ( \sqrt{\sum_{k=1}^{m}(x_{ik} - w_{jk})^2} \right ).
\end{equation}
The weights of the winning neuron and its neighboring neurons (determined by a neighborhood radius function) are then updated to make them closer to the training vector.

\subsection{MECshield Framework}\label{MECshield Framework}
Figure \ref{fig:architecture} illustrates the proposed MECshield framework. Logically, the MECshield framework consists of a central controller and multiple agents located at the edge of each local network. The communication between the central controller and the distributed agents is facilitated via secure channels/protocols supported by the H-IoT networks (e.g., Openflow protocol on secure management channels in SDNFV technology).

\textbf{MECshield controller:} The main purpose of the central controller is to orchestrate operations among the distributed agents. A local network might act as the source or destination of DDoS attacks. Therefore, each local network requires a different smart SOM filtering strategy, based on its position in the attack scenario. For instance, an attack-destination site that suffers extreme flooding traffic should deploy an appropriate SOM training strategy to focus on the traffic \textit{protocol}, \textit{port number}, and \textit{flow number}, including the IP address classes of the incoming traffic. Alternatively, the source site(s) of the attack should be more concerned with features of its outgoing traffic for individual sources, such as the \textit{protocol}, \textit{port number}, \textit{flow number}, \textit{packet number} per flow, and \textit{transmission contiguity}. Thus, only the attack-source sites participate in DDoS defense. Therefore, cooperation among the distributed local agents is crucial for an effective defense. The components of the MECshield controller are described as follows:

\begin{itemize}
    \item \textit{Report collector:} This component gathers traffic reports from distributed agents including traffic protocols, port ranges, volume or traffic flow quantity, source IP address ranges, and destination IP address(es). Adopting the requirements of the detection mechanism applied in the attack analyzer, the collected information is pre-processed and updated at the report collector before being transferred to the attack analyzer. For instance, to analyze the characteristics of a spoofed DDoS flooding attack, protocol, port range and volume are necessary for attack investigation.
    
    \item \textit{Attack analyzer:} First, DDoS attack detection techniques \cite{6489876} are utilized to identify the attack symptom based on the processed information from the report collector. Once the attack is identified, further information is then considered to recognize attack targets, attack methods. For example, Smurf and fraggle attacks produce tremendous numbers of packets on some flows to exhaust the capability of the victim, while TCP SYN attacks establish large numbers of connections to the victim during a short period \cite{phan2017distributed}.
    
    \item \textit{Policy generator:} Once the DDoS attack is identified, primary policies are generated and forwarded to the MECshield agents located at the edges of the source and destination sites of the attack. The policies contain summary information of the attack (target, attack method and possible mitigation policies) and desired features. The feature information will be delivered to the feature extractor module via the local policy conductor at both source and destination site agents in order to request for desired extraction, which is used in the SOM classification process.
\end{itemize}

\textbf{MECshield agent:} The primary purpose of MECshield agents is to mitigate the DDoS traffic and the components of MECshield agent are as follows:

\begin{itemize}
    \item \textit{Traffic monitor:} The main function supported by this component is to make the traffic statistics report. It regularly captures statistics of incoming traffic, including traffic protocols, service ports, volumes and source/destination IP addresses. This information is delivered to the report collector in the central controller. In addition, incoming traffic is also forwarded to the feature extractor in order to make the SOM map's inputs. 
    
    \item \textit{Local policy conductor:} Based on the primary policy dispatched from the controller, the local policy conductor informs the feature extractor about prominent features in order to make a tuple of features for each input vector in the SOM map classification procedure. Moreover, the local policy conductor will send mitigation information to the smart SOM filter agent to apply appropriate policies for attack traffic. For example, a drop action should be given to TCP SYN flooding attack flows because of the number of flow is huge and the packet per flow is tiny, meanwhile a blocking action should be applied for attack flows transferring a large amount of packet in a flow.

    \item \textit{Feature extractor:} This component extracts the features of traffic delivered from the traffic monitor and make tuples for the SOM inputs based on requirements of the local policy conductor. Then, it forwards these tuples to the smart filter agent for classifying and training.
    
    \item \textit{Smart SOM filter:} First, the SOM is trained continuously by input vectors transferred from the feature extractor. Second, when a vector of DDoS attack traffic is recognized at the SOM, the smart SOM filter notifies the switching/routing component. Consequently, a protection mode is activated and the outgoing traffic is switched through the filter. The protection mode is deactivated if the SOM does not receive a training vector of a DDoS attack within a pre-defined duration. Finally, mitigating the DDoS attack by dropping the traffic is classified by the SOM as malicious.
    
    \item \textit{Switching/Routing:} This is a basic function of edges used to handle incoming/outgoing traffic.
\end{itemize}


\subsection{Operational Workflow}
To describe the operational workflow of MECshield, we consider a normal and a DDoS attack.

Under normal conditions, the protection mode is deactivated; that is, outgoing traffic from IoT devices bypasses the smart SOM filter to improve the networking performance. In this case, the traffic is still captured by the traffic monitor to extract features for smart SOM training and for traffic statistics reports (yellow lines in Figure \ref{fig:architecture}). Whenever a DDoS symptom is detected by the attack analyzer in the controller or by the smart SOM filter in the local agent, the protection mode is activated. 

In the DDoS attack condition, the outgoing traffic from IoT devices should go through the smart SOM filter. Depending on the classification provided by the filter, detected DDoS traffic is dropped. The traffic monitor collects statistics on the DDoS traffic, which it reports to the controller. After identifying the attack targets and attack methods, the controller dispatches primary policies to all agents distributed at the edge of the corresponding local networks. The local policy conductor in each MECshield agent assigns requirements to the feature extractor to generate appropriate training vectors, and it also informs the smart SOM filter about possible mitigation policies to tackle attack traffic flows. During attack time, the smart SOM filter still transfers incoming attack traffic (red lines in Figure \ref{fig:architecture}) to the traffic monitor and the feature extractor to generate statistics and make training samples, respectively.

The security performance of the proposed framework is achieved through two layers of protection given by the MECshield agents at the attack-source and destination sites. 

\section{Performance Evaluation}
This section describes the experiment setup, and then provide an in-depth performance analysis of the DDoS defense system.    

\subsection{Experiments}
\subsubsection{Training data sets for the smart SOM filter}
Initially, the smart SOM filters are trained using data sets of DDoS attack and normal traffic. The DDoS-attack training sets are obtained from three data sets: CAIDA-attack-traffic \cite{CAIDA}, NSL-KDD \cite{NSL-KDD}, and DARPA 2009 Intrusion Detection \cite{DARPA2009}. The normal-traffic training set is derived from CAIDA-normal-traffic \cite{CAIDA}. Statistics of these data sets are provided in Table \ref{tab:SOMTrainingDatasets}.
\begin{table}[!t]
\centering
\caption{Statistical Information of CAIDA, NSL-KDD and DARPA Data Sets}
\label{tab:SOMTrainingDatasets}       
\begin{tabular}{cccc}
\hline
\textbf{CAIDA}\\
\hline
Traffic state&TCP&ICMP&Others\\
&(\%)& (\%)& (\%)\\
\hline
Normal (2015)&88.45&6.0&5.55\\
\hline
Attack (2007)&7.58&91.25&1.17\\
\hline
\textbf{NSL-KDD}\\
\hline
Attack &Training&Testing&Features\\
types&patterns&patterns&\\
\hline
back, land, neptune,&45927&7458&41\\
pod, smurf, teardrop\\
\hline
\textbf{DARPA}\\
\hline
Attack types& Attack source& Attack time&\\
\hline
SYN flooding& 100 different IPs & 6 minutes&\\
\hline
\end{tabular}
\end{table}

Owing to the wide variety of H-IoT devices, we generalize the types of traffic into three categories:
\begin{itemize}
\item \textit{Sensor traffic}: This traffic is generated by sensor devices in a fixed period, with a low number of packets per flow.
\item \textit{Monitor traffic}: This involves real-time traffic, characterized by a small number of flows and a significant number of packets per flow (e.g., camera). 
\item \textit{Alarm traffic}: This traffic type is not easily discernible because alarm IoT devices only generate traffic when an abnormal event occurs. However, we assume the alarm traffic has both moderate flows and a moderate number of packets per flow.
\end{itemize}

Therefore, only samples that belong to the three categories mentioned above are extracted from the CAIDA, NSL-KDD, and DARPA data sets to train the SOM filters. Accordingly, a tuple (\textit{protocol}, \textit{port number}, \textit{flow number}) is applied for SOM training in the MECShield agent at the destination-site; at the source-site, a tuple (\textit{protocol}, \textit{port number}, \textit{flow number}, \textit{packet/flow}, \textit{transmission contiguity}) is used. Then, each traffic type is trained by a set of 10000 samples. And, key parameters of the SOM map are as follows: neuron number is 400, learning rate is set to 0.1, and output dimension is 2.

\subsubsection{Emulation Setup}
To emulate a distributed MEC network, we set up a SDNFV-enabled network of four physical servers. This comprises the MECshield controller and three MECshield agents directly connected to three IoT networks (sensor, monitor, and alarm) comprising both SOM filtering and switching/routing features. For convenience, we implement the MECshield agent as a software-based box including SOM, OpenvSwitch functionality and other modules. Furthermore, we can configure the OpenvSwitch agent to forward local traffic to a smart SOM filter before leaving the machine \cite{phan2017distributed}. Applications in MEC servers store IoT incoming traffic and send back to IoT clients a simple response message confirming receipt of the data.

\subsubsection{Testing Process}
To assess the proposed framework, a test is carried out, along with two other schemes (a \textit{Centralized}-SOM and a \textit{Distributed}-SOM \cite{phan2017distributed}). In the \textit{Centralized}-SOM solution, a SOM filter is placed at the controller, and all H-IoT traffic is forwarded to the controller for analysis. Meanwhile, in the \textit{Distributed}-SOM mechanism, the SOM maps trained by all agents are merged at the controller after the training period. Afterward, the merged SOM is returned to the agents for local traffic handling \cite{phan2017distributed}. Note that all solutions use the same training data sets. For each local IoT network, we use the \textit{BoNeSi simulator} tool to generate different levels of attack traffic (50, 100, 200, and 300 Mbps) in all schemes.

\subsection{Emulation Results and Analysis}

\begin{figure}
\centering
\includegraphics[width=0.48\textwidth]{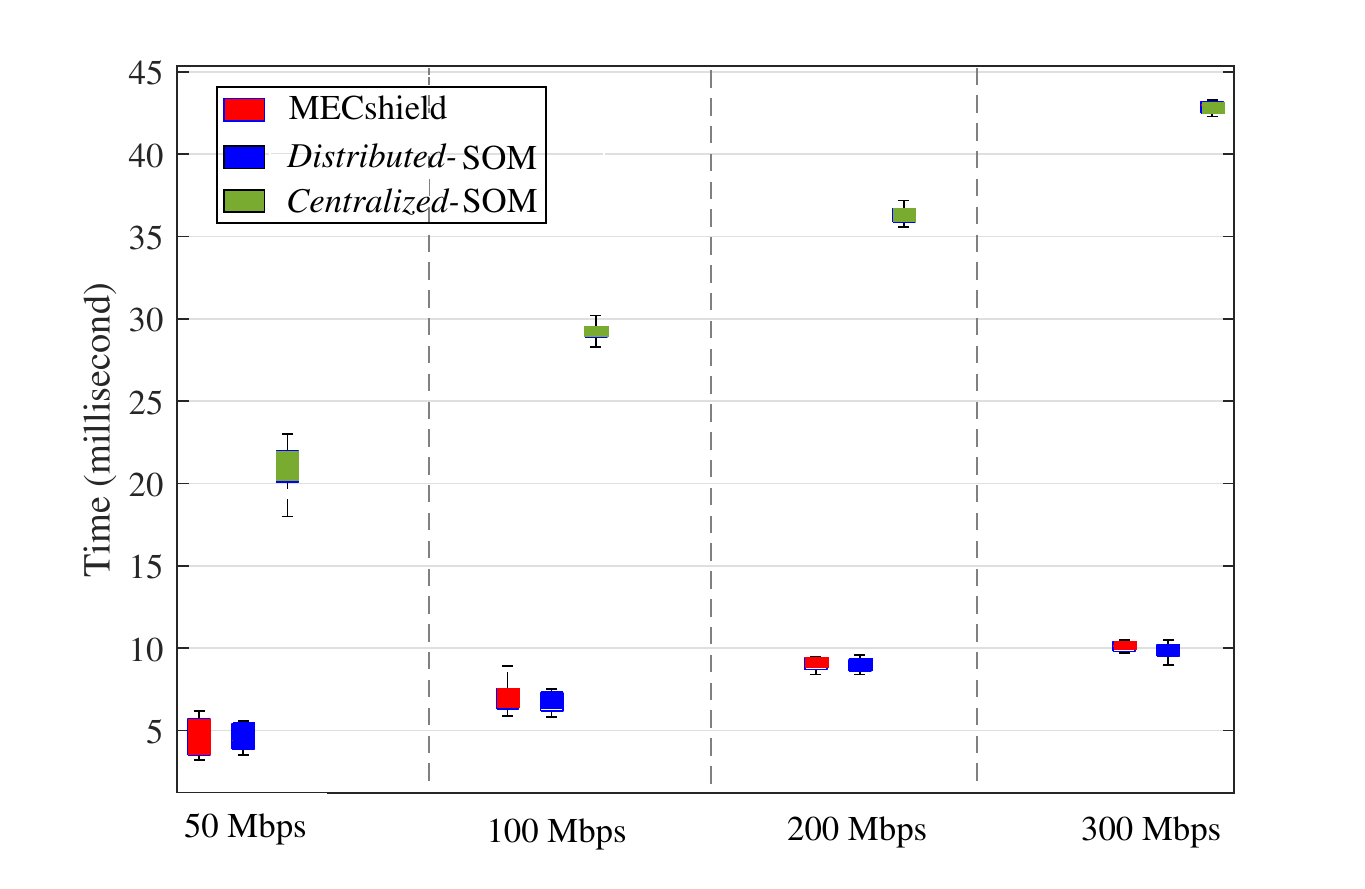}
\caption{Reaction time to various attack levels}
\label{fig:AttackReaction}
\end{figure}

\begin{figure}
\centering
\includegraphics[width=0.48\textwidth]{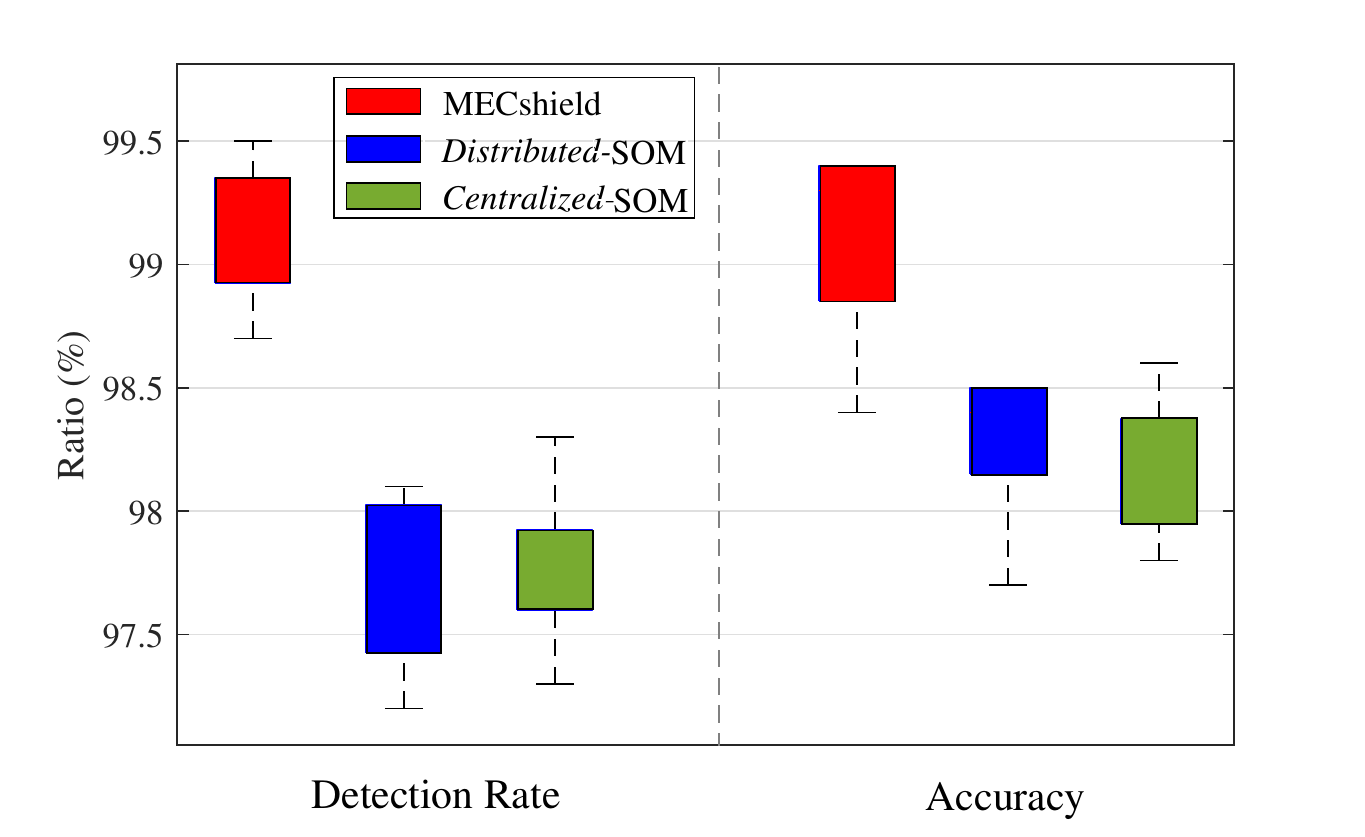}
\caption{Detection Rate and Accuracy in classifying abnormal traffic with the SOM map}
\label{fig:DRACC}
\end{figure}

\begin{figure}
\centering
\includegraphics[width=0.48\textwidth]{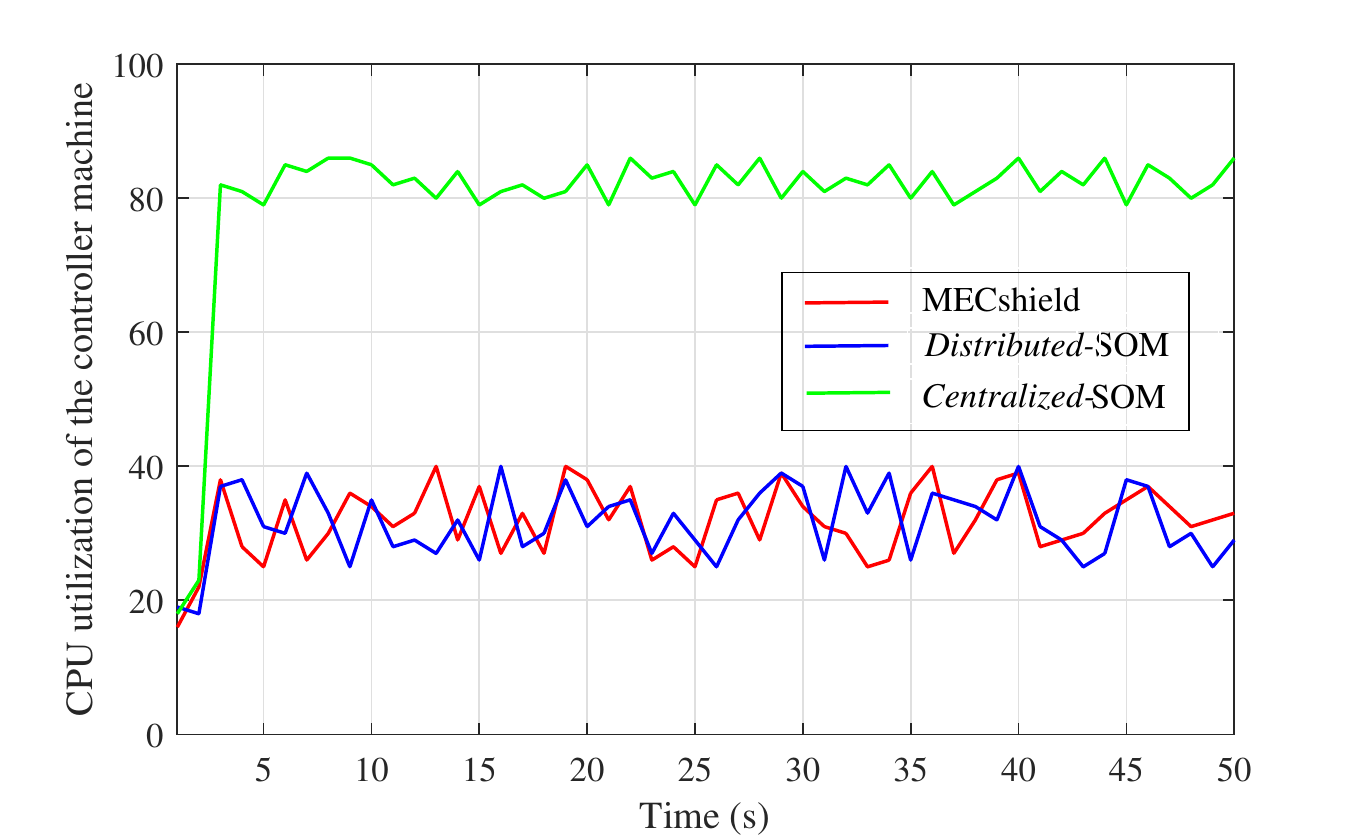}
\caption{CPU utilization in controllers under DDoS attacks}
\label{fig:Performance}
\end{figure}

\subsubsection{Attack Reaction Performance}
The first criterion is the attack reaction time at each edge agent. In this measurement, we record four different attack traffic levels, as depicted in Figure \ref{fig:AttackReaction}. These results can be explained as follows:
\begin{itemize}
\item In the \textit{Centralized}-SOM, there is a considerable delay because traffic is forwarded to the controller from the edge devices for attack investigation. Afterwards, policies are sent back to edges for appropriate traffic-handling operations.

\item In the other schemes, if the smart SOM filter at an edge agent detects malicious patterns, it immediately applies the determined policies to the attack traffic by preventing it from entering the distribution network. Therefore, the time taken to react to attack patterns of the MECshield framework and the \textit{Distributed}-SOM solution are lower for all traffic levels.
\end{itemize}

\subsubsection{Detection rate and Accuracy Improvements}
As a second criterion, we measure the detection rate and accuracy of three schemes. Figure \ref{fig:DRACC} presents the detection rate and accuracy of the three approaches. In both criteria, the MECshield performed better than the other schemes. This is because SOM maps in the MECshield agents are separately trained by different local IoT traffic. Hence, these filters find it easier to recognize patterns or they are well-adapted to local IoT traffic in other word. Conversely, with a fixed and limited number of neurons in a smart SOM agent, if there are many traffic types trained for a SOM map (\textit{Centralized}-SOM case), or several merging times in the case of a \textit{Distributed}-SOM mechanism, the weights of each neuron in the SOM map will change considerably. This leads to the degradation of both the detection rate and the accuracy of these schemes.

\subsubsection{Bottleneck-Handling Performance}
To assess the robustness of the schemes, we investigate the problem of bottlenecks occurring in the controller during our experiments. The results are shown in Figure \ref{fig:Performance}. A major difference is observed between distributed and centralized solutions. Both MECshield and the \textit{Distribued}-SOM show acceptable CPU usage of around 35\%. However, the \textit{Centralized}-SOM mechanism shows a high usage of the controller's CPU (83\%, on average); this is because edge node traffic is always forwarded to the controller for processing, which becomes the bottleneck during DDoS attacks.

\subsubsection{Overall CPU Resource Consumption}
Finally, we assess the MECshield framework's overall CPU resource consumption in the case of DDoS attacks coming from several IoT networks. We record the CPU usage of all machines and evaluate the average system resource consumption. The CPU usage of MECshield, the \textit{Distributed}-SOM, and the \textit{Centralized}-SOM are 36\%, 43\%, and 46\%, respectively. As discussed in Section \ref{MECshield Framework}, we consider the IP ranges of incoming traffic. Therefore, depending on the IP ranges, the MECshield controller can inform dedicated agents to enable the SOM filter function in the case of attacks. As a result, the MECshield framework can save resources because of the limited number of running SOM filters. In contrast, the \textit{Distributed}-SOM and \textit{Centralized}-SOM schemes always have to enable all SOM filter agents. Hence, the computing resources are consumed, even if there is no incoming traffic.

\section{Conclusion}
In this study, we propose MECshield, a DDoS prevention framework that leverages MEC power to deploy multiple smart filters at the edge of attack-source/destination networks. MECshield enables the network to defend against malicious traffic from H-IoT devices through smart SOM filters deployed in front of the attack source. Experimental results show that the detection rate and accuracy are improved because of the well-adaptation to local traffic at the SOM filters. Moreover, the distributed architecture and control scheme of the MECshield avoids the bottleneck occurring in DDoS attacks, and saves around 10\% resource consumption in terms of CPU usage compared to other methods. Finally, MECshield introduces an efficient and feasible security framework for an H-IoT environment.

\bibliographystyle{IEEEtran}

\begin{thebibliography}{10}
\bibitem{gartner2017report}
{Gartner, Inc.}, ``Gartner says 8.4 billion connected "things" will be in use
  in 2017, up 31 percent from 2016,'' {(cited Oct. 10, 2017)}. [Online].
  Available:http://www.gartner.com/newsroom/id/3598917
\bibitem{kolias2017ddos}
C.~Kolias, G.~Kambourakis, A.~Stavrou, and J.~Voas, ``{DDoS in the IoT: Mirai
  and other botnets},'' \emph{IEEE Computer}, vol.~50, no.~7, pp. 80--84, 2017.
\bibitem{sun2016edgeiot}
X.~Sun and N.~Ansari, ``{EdgeIoT}: Mobile edge computing for the {I}nternet of
  things,'' \emph{IEEE Communications Magazine}, vol.~54, no.~12, pp. 22--29,
  2016.
\bibitem{kohonen2013essentials}
T.~Kohonen, ``Essentials of the self-organizing map,'' \emph{Neural networks},
  vol.~37, pp. 52--65, 2013.
\bibitem{lin2017survey}
J.~Lin, W.~Yu, N.~Zhang, X.~Yang, H.~Zhang, and W.~Zhao, ``A survey on
  {I}nternet of things: architecture, enabling technologies, security and
  privacy, and applications,'' \emph{IEEE Internet of Things Journal}, vol.~4,
  no.~5, pp. 1125--1142, 2017.

\bibitem{alam2017data}
F.~Alam, R.~Mehmood, I.~Katib, N.~Albogami, and A.~Albeshri, ``Data fusion and
  {IoT} for smart ubiquitous environments: A survey,'' \emph{IEEE Access},
  vol.~5, pp. 9533--9554, 2017.
\bibitem{hu2015mobile}
Y.~C. Hu, M.~Patel, D.~Sabella, N.~Sprecher, and V.~Young, ``Mobile edge
  computing -- {A} key technology towards {5G},'' \emph{ETSI White Paper No.
  11}, 2015.
\bibitem{aleroud2017contextual}
A.~Aleroud and G.~Karabatis, ``Contextual information fusion for intrusion
  detection: a survey and taxonomy,'' \emph{Knowledge and Information Systems},
  vol.~52, no.~3, pp. 563--619, 2017.
\bibitem{network2017worldwide}
D.~Anstee, P.~Bowen, C.~F. Chui, and G.~Sockrider, ``Worldwide infrastructure
  security report,'' \emph{{Arbort Networks special report -- Volume XII}},
  2017.
\bibitem{leverett2017towards}
E.~Leverett and A.~Kaplan, ``Towards estimating the untapped potential: a
  global malicious {DDoS} mean capacity estimate,'' \emph{Journal of Cyber
  Policy}, vol.~2, no.~2, pp. 195--208, 2017.
\bibitem{6489876}
S.~T. Zargar, J.~Joshi, and D.~Tipper, ``A survey of defense mechanisms against
  distributed denial of service (ddos) flooding attacks,'' \emph{IEEE
  Communications Surveys Tutorials}, vol.~15, no.~4, pp. 2046--2069, Fourth
  2013.
\bibitem{phan2017distributed}
T.~V. Phan, N.~K. Bao, and M.~Park, ``{Distributed-SOM}: A novel performance
  bottleneck handler for large-sized software-defined networks under flooding
  attacks,'' \emph{Journal of Network and Computer Applications}, vol.~91, pp.
  14--25, 2017.
\bibitem{CAIDA}
{CAIDA}, ``{The CAIDA datasets of anonymized Internet traces and DDoS
  attack},'' {(cited Sep. 10, 2017)}. [Online]. Available: https://data.caida.org/datasets/
\bibitem{NSL-KDD}
{NSL-KDD}, ``Data set for network-based intrusion detection systems,'' {(cited
  Sep. 10, 2017)}. [Online]. Available:https://data.caida.org/datasets/security/ddos-20070804/
\bibitem{DARPA2009}
{LANDER}, ``{LANDER:DARPA DDoS attack-20091105},'' {(cited Sep. 10, 2017)}.
  [Online]. Available:https://ant.isi.edu/datasets/readmes/DARPA-2009-DDoS-attack-20091105.README.txt
\end{thebibliography}

\end{document}